\documentclass[11pt,twoside]{article}
\usepackage{asp2004}
\usepackage{psfig}
\usepackage{epsf}
\usepackage{floatflt,graphicx}
\usepackage{lscape}
\def\edcomment#1{\iffalse\marginpar{\raggedright\sl#1\/}\else\relax\fi}
\reversemarginpar

\begin{document}
\title{Steep functions in astronomy: the RQSO $z$-cutoff debate}
 \author{J. V. Wall}
\affil{Department of Physics and Astronomy, University of British
Columbia, 6224 Agricultural Road, B.C. V6T 1Z1, Canada}

\begin{abstract}
 The debate over the
existence of a redshift cutoff for radio-loud QSOs (RQSOs) hinges
on interpreting a second-order effect -- radio spectral variations
in variable QSOs initially selected by survey from a population
with a steep source count. I discuss the resolution of this
question; the issue is highly relevant to modern surveys and
follow-up observations.
\end{abstract}

\vspace{-0.5cm}
Astronomers find steep, open-ended functions difficult: the
log~$N$~–-~log~$S$ curve for example, or the
probability-area-radius relation in making cross-wave\-band
identifications. The argument (Jarvis and Rawlings 2000; Wall et
al. 2005) over a RQSO redshift cutoff is an instance in which a
{\it second-order effect} from steep selection functions causes
discrepant results in different analyses.
\begin{floatingfigure}[l]{5.5cm}
      \centering
      \includegraphics[scale=0.27]{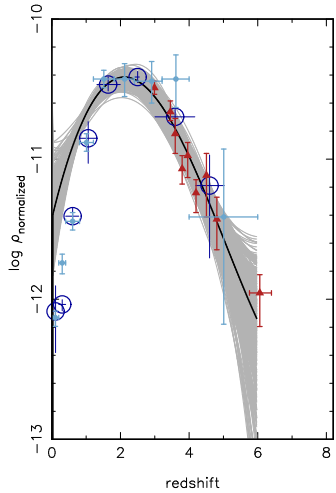}
      \caption{\footnotesize{Space densities of optical-, X-ray- and radio-selected QSOs.}}
      \label{zcutoff}
    \end{floatingfigure}
Such a cutoff is present in optically-sel\-ected QSOs of the SDSS
(Fig.~\ref{zcutoff}, triangles) and in X-ray selected QSOs of XMM,
ROSAT and CHANDRA (Fig. 1, circles, points). A similar cutoff for
radio-loud QSOs is important because if present it points to a
real cutoff; obscuration cannot be responsible. Such a cutoff then
defines an epoch of creation for galaxy hosts of massive black
holes powering AGN, a datum for e.g.\ hierarchical galaxy
formation in a CDM universe. Jarvis and Rawlings (2000) concluded
that the presence of an RQSO redshift cutoff was unproven. In
contrast, using a similar sample of RQSOs Wall et al (2005) found
conclusive evidence for a decline of space density to high
redshifts (Fig.~\ref{zcutoff}, curve + shaded lines).

Testing for a RQSO redshift cutoff must start with dogged hard
work to produce a sample of radio-loud QSOs complete to a radio
flux-density limit, and with $\sim$ complete optical
identification data, including redshifts. The Parkes 2.7-GHz
quarter-Jansky sample (Jackson et al. 2002) fits the bill: it
covers much of the southern sky, and contains a sub-sample of
$\sim 400$ QSOs with spectra selected to have spectral indices
$\alpha\ (\rm{with}~ S\propto\nu^\alpha)
>-0.4$ in the range 2.7 to 5.0 GHz. Most of these RQSOs have
redshift determinations.

A statistical method to examine variation in space density is to
take the $1/V$ space-density contributions for all QSOs over some
epoch, say $1 < z < 3$, and use these to predict space density
contributions at epochs out to which the survey permits them to be
seen, e.g. $3 < z < 8$. The objects become invisible at the larger
redshifts because either flux density falls below the survey
limit, or the spectral index in the observer range (2.7 - 5 GHz)
falls below $-0.4$. The contributions of each source at higher
redshifts are then added, and compared with the number of RQSOs
observed. Two related issues in this approach are:

(1) the detection-limit line in the luminosity-redshift plane for
the radio survey. This differs for every object because the
GHz-range radio spectra of RQSOs scatter wildly, with single or
multiple `bumps', and/or steep drop-offs or steep rises. The issue
was resolved by the invention of the `single-source-survey'
technique (Wall et al. 2005), in which every source is given its
own luminosity-redshift plane, with a `survey' cutoff line unique
to its radio spectrum.
\begin{floatingfigure}[r]{5.5cm}
      \includegraphics[scale=0.25]{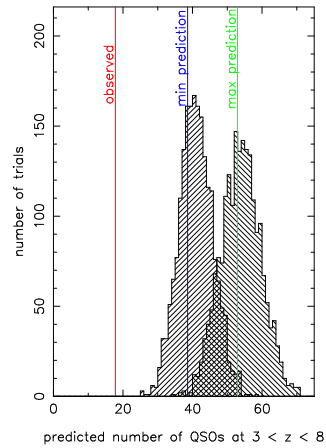}
      \caption{\footnotesize{Bootstrap
      predictions for uniform space
      density of RQSOs at $z > 3$.}}
      \label{boot}
    \end{floatingfigure}
(2) spectral measurement, in the presence of variability. If flux
measurements at frequencies above the survey frequency are biased
low, then spectra appear to steepen, objects can be `seen' to
smaller distances, and a low estimate of expected numbers at high
redshifts follows. Most high-frequency measurements for our sample
and for the sample of Jarvis and Rawlings (2000) were made at 8.4
GHz, {\it $~$25 years after} the initial finding survey. But all
radio surveys primarily select variable\- sources biased high,
with survey-frequency flux densities above that of the
time-average for each source. This is from the steep source count:
simulations show that for a count of slope $-2.5$, 78\% of all
variable sources are selected in the up-state. Higher-frequency
measurements made years later result in a time-average of the flux
density; this in combination with survey data means that the
average high-frequency spectrum is biased to be too steep.

In fact we were lucky to have 8.9-GHz measurements of the 40
brightest objects in the sample {\it taken at the time of the
2.7-GHz survey} (Shimmins and Wall 1973). Because the brightest
RQSOs make the major contribution to the numbers predicted at
$z>3$, this was just enough data to correct for the bias;
furthermore, it demonstrated the extent of the bias. In 25 years,
8-GHz intensities had changed by factors of up to 4.
Fig.~\ref{boot} shows end-to-end bootstrap tests of our analysis.
Assuming uniform space distribution, all of the 4000 predictions
lie well above the total of 18 $z>3$ QSOs observed in our sample.
The relative dearth of RQSOs beyond $z>3$ is highly significant.

{\it If variability is involved, intensities (line or continuum)
measured after the finding survey will inevitably bias the
spectrum of the sample.} It is not obvious that a post-hoc
statistical correction can be applied. This has serious
implications for all modern multi-wavelength studies in which
variability is involved; biases are inevitable and analyses
ignoring them are suspect.

\begin{footnotesize}
REFERENCES: Jackson, C.A., et al. 2002, Astron. \& Astrophys. 386,
160; Jarvis, M.J. and Rawlings, S. 2000, Mon. Not. R. Astr. Soc.
319,121; Shimmins, A.J. and Wall, J.V. 1973, Austr. J. Phys. 26,
93; Wall, J.V., et al. 2005, Astron. \& Astrophys. 434, 133.
\end{footnotesize}

\end{document}